\documentclass[preprint,11pt,authoryear]{elsarticle}
\usepackage{geometry}
\usepackage{amssymb}
\usepackage{graphicx} 
\usepackage{amsthm}
\usepackage{amsmath}
\usepackage{lineno}
\usepackage{hyperref}
\usepackage{tabularx}
\usepackage{url}
\usepackage{makecell}
\setcitestyle{numbers,square}
\usepackage{bm}
\usepackage{amsmath}
\usepackage{amsfonts}
\usepackage[dvipsnames]{xcolor}
\usepackage{float}
\usepackage{xr}
\usepackage{framed}
\usepackage{makecell}
\usepackage{arydshln}
\usepackage{fancyhdr}
\usepackage{booktabs}
\usepackage{fullpage}
\usepackage{enumitem}
\usepackage{subcaption}

\usepackage[linesnumbered,boxed,ruled,vlined,longend]{algorithm2e}

\definecolor{darkgreen}{rgb}{0.8,.4,0.4}

\journal{arxiv.org}
\begin{document}
\begin{frontmatter}
\title{\textbf{\Large Stop-and-go wave super-resolution reconstruction\\ via iterative refinement}}
\author[inst1,inst2]{Junyi Ji}
\ead{junyi.ji@vanderbilt.edu}
\author[inst2,inst3]{Alex Richardson}
\author[inst2]{Derek Gloudemans}
\author[inst2]{Gergely Zach\'ar}
\author[inst2]{Matthew Nice}
\author[inst2]{William Barbour}
\author[inst1,inst2,inst3]{Jonathan Sprinkle}
\author[inst4]{Benedetto Piccoli}
\author[inst1,inst2,inst3]{Daniel B. Work}

\affiliation[inst1]{organization={Department of Civil and Environmental Engineering, Vanderbilt University},country={United States}}
\affiliation[inst2]{organization={Institute for Software Integrated Systems, Vanderbilt University},country={United States}}
\affiliation[inst3]{organization={Department of Computer Science, Vanderbilt University},country={United States}}
\affiliation[inst4]{organization={Department of Mathematical Sciences, Rutgers University–Camden},country={United States}}
\begin{abstract}
Stop-and-go waves are a fundamental phenomenon in freeway traffic flow, contributing to inefficiencies, crashes, and emissions. Recent advancements in high-fidelity sensor technologies have improved the ability to capture detailed traffic dynamics, yet such systems remain scarce and costly. In contrast, conventional traffic sensors are widely deployed but suffer from relatively coarse-grain data resolution, potentially impeding accurate analysis of stop-and-go waves. This article explores whether generative AI models can enhance the resolution of conventional traffic sensor to approximate the quality of high-fidelity observations. We present a novel approach using a conditional diffusion denoising model, designed to reconstruct fine-grained traffic speed field from radar-based conventional sensors via iterative refinement. We introduce a new dataset, \texttt{WaveX}, comprising 132 hours of data from both low and high-fidelity sensor systems, totaling over 2 million vehicle miles traveled. Our approach leverages this dataset to formulate the traffic measurement  refinement problem as a spatio-temporal super-resolution task. We demonstrate that our model can effectively reproduce the patterns of stop-and-go waves, achieving high accuracy in capturing these critical traffic dynamics. Our results show promising advancements in traffic data refinement, offering a cost-effective way to leverage existing low spatio-temporal resolution sensor networks for improved traffic analysis and management. We also open-source our dataset, trained model and code to enable further research and applications.
\end{abstract}
\end{frontmatter}

\section{Introduction}

Stop-and-go waves are among the most ubiquitous and significant phenomena in freeway traffic flow research\cite{ahn2019traffic,edie1961,edie1967generation}. Stop-and-go waves have been recognized as a major contributor to the system inefficiencies, crashes \cite{zheng2010impact} and emissions \cite{treiber2008much,li2014stop}. Modern freeway \textit{traffic management centers} (TMCs) commonly utilize sensors such as loop detectors \cite{chen2002freeway} and radar detectors \cite{kim2017assessing} in daily operations to implement various corridor management strategies \cite{siri2021freeway}. These sensors typically provide aggregated traffic measurements at typical intervals of 30 seconds to 5 minutes and are usually deployed at 0.5 to 1-mile intervals along the freeway. Because of such sensors, we know that stop-and-go traffic waves are extremely prevalent in freeway corridors \cite{treiber2000congested,he2025review}. However, while radar and inductive loop sensors offer a good cost-benefit ratio compared to expensive high-fidelity sensors for TMCs, their lack of fine spatio-temporal resolution means that they cannot accurately measure periods of low-speed travel temporally close to periods of high-speed travel \cite{coifman2001improved,kim2014comparing}. This underestimation is particularly problematic as these low-speed parts are critical for accurately estimating stop-and-go wave characteristics. 

\begin{figure}[H]
    \centering
    \includegraphics[width=0.82\linewidth]{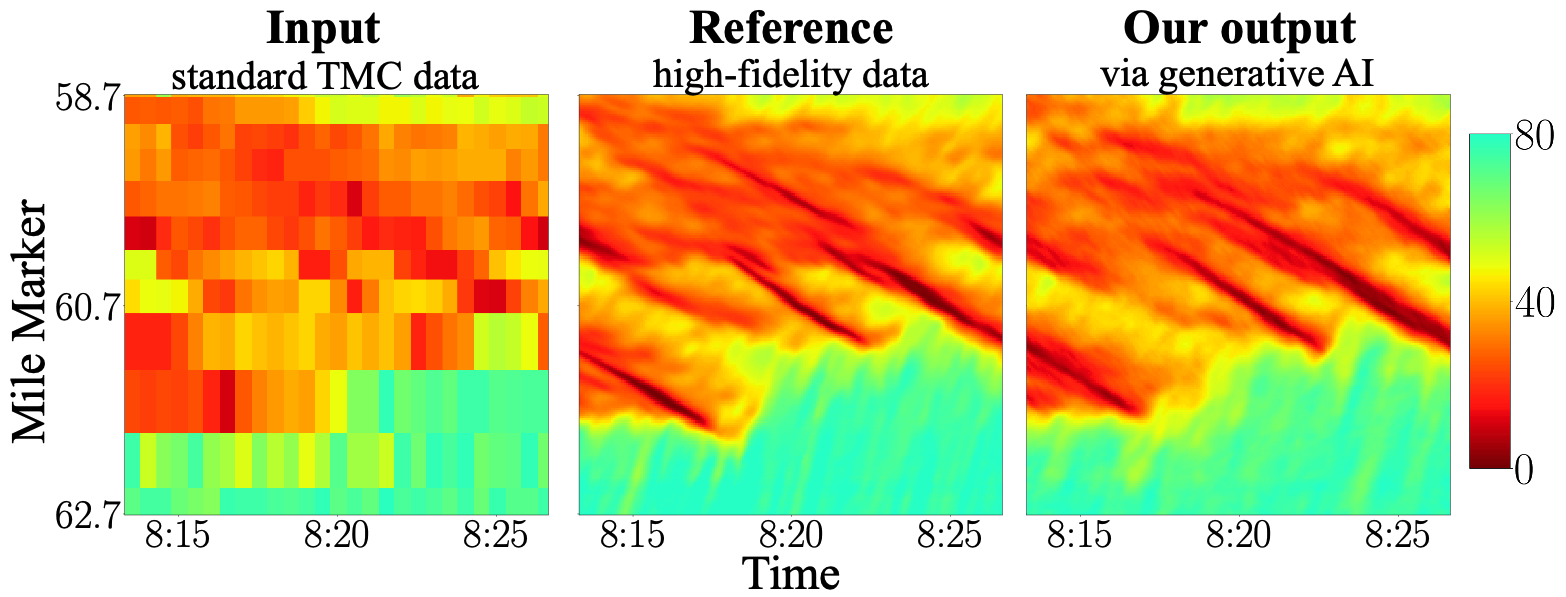}
    \caption{Stop-and-go waves observed from a standard traffic sensor system (left), a high-fidelity sensor system (middle), and a generative model output (right) using the low spatio-temporal resolution data as input. The examples used here are extracted from Interstate 24 near Nashville, Tennessee in the United States, dated June 3rd, 2024 during the morning peak hours. In these examples, the x-axis represents the local time, and the y-axis represents the mile marker of the freeway. The high-fidelity sensors distinctly highlight the stop-and-go wave patterns, providing a much clearer and detailed view of the traffic flow dynamics.} 
    \label{fig:overview}
\end{figure}

In recent years, advanced sensor technologies such as drones \cite{barmpounakis2020new,talebpour2024}, multi-camera monitoring systems \cite{shi2021video, gloudemans202324}, LiDAR \cite{lee2015using,coifman2016segmenting}, and on-board vehicle perception systems \cite{sun2020scalability} have allowed the collection of precise vehicle trajectories, which has enabled rapid advancement in the analysis and modeling of stop-and-go waves \cite{treiterer1974hysteresis, NGSIM, ztd2018, gloudemans202324}. High-fidelity traffic measurements have demonstrated the capability to capture the intricate patterns of stop-and-go waves, which exhibit considerable complexity, as shown in \cite{laval2010mechanism, ji2024enabling}. Figure \ref{fig:overview} demonstrates a comparison of the traffic speed profiles observed from a conventional sensor system and high-fidelity sensor systems.  Unfortunately, very few roadways worldwide currently have access to high-fidelity traffic instruments mentioned above capable of continuously collecting extensive vehicle trajectory data \cite{li2020trajectory}. This significantly limits the generalizability and scalability of the observation and analysis of stop-and-go traffic to a broader context. 

Recent years have seen a flurry of advancements in generative artificial intelligence (AI) models, and their diverse applications \cite{brown2020language,ho2020denoising} have leveraged the potential of large-scale data via generative models, motivating the primary thesis of this work: \textbf{generative models can learn the underlying mechanisms of traffic from low spatio-temporal resolution sensor input to generate enhanced data which better capturing stop-and-go waves, almost to the resolution of high-fidelity traffic sensors.}

To investigate this, we collected 132 hours of rush-hour traffic data over 50 days from a stretch of 4 miles of roadway, amounting to over 2 million vehicle miles traveled (VMT). The dataset, named \texttt{WaveX}, was gathered simultaneously from two distinct sensor systems: one from a commonly used freeway radar detector system (with sensors deployed every 0.3 to 0.5 mile and aggregated every 30 seconds), and the other from a camera-based traffic testbed capable of capturing individual vehicle trajectories (aggregated into speed profile data with bin resolution of 4 seconds and 0.02 miles, approximately 32 meters) \cite{gloudemans202324,ji2024virtual}. We then formulated the data refinement problem as a traffic measurement super-resolution reconstruction task, with the goal of reconstructing the higher-fidelity data using only the lower fidelity data as input. To this end, we implemented a conditional diffusion denoising model for iterative refinement. Finally, we evaluated the extent to which our proposed method accurately reconstructs the details of stop-and-go waves.

The contributions of this paper are:
\begin{enumerate}[label=(\roman*),noitemsep]
\item We are the first to apply a generative conditional diffusion denoising model designed to improve the spatio-temporal resolution of conventional freeway roadside traffic sensors, addressing the traffic measurement super-resolution challenge.
\item We produce the \texttt{WaveX} dataset, a new large-scale empirical dataset over 2 million vehicle miles traveled featuring stop-and-go traffic, which was collected concurrently using both high-resolution and low-resolution sensor systems over 50 days, 4 hours a day. This dataset is intended to investigate the potential of generative models in analyzing traffic flow. 
\item We demonstrate that our model can reproduce wave properties, effectively capturing the complex details of stop-and-go traffic patterns and significantly outperforming existing methods across a variety of metrics. 
\item We release the dataset along with the code for community-wide use to enable future
exploration of generative AI in transportation and support the
reproducible research. 
\end{enumerate}

The remainder of this article is structured as follows: Section~\ref{sec:related} provides a brief overview of related work on traffic waves and the use of generative models in traffic. Section~\ref{sec:method} details the proposed method. Section~\ref{sec:data} presents the collected data, while the experiments conducted to validate our approach are described in Section~\ref{sec:exp}. The results are outlined in Section~\ref{sec:result}, and further discussed in detail in Section~\ref{sec:discussion}. Section~\ref{sec:conclusion} highlights the ability of our method to capture fine-grained wave properties and discusses its limitations.

\section{Related work and challenges}
\label{sec:related}
Traffic reconstruction across all scales \cite{piccoli2015second} requires data collection as the first step. Efforts in empirical traffic data collection have been ongoing since 1933 \cite{greenshields1933photographic}, and these efforts can be categorized into \textit{microscopic} and \textit{macroscopic} scales.
Macroscopic traffic data \cite{greenshields1933photographic, cassidy1999some, daganzo1999possible, treiber2000congested, coifman2002estimating, kim2017assessing}, which refers to the aggregated traffic measurements typically collected from loop detectors, provides insights into collective phenomena such as the fundamental diagram and capacity drop \cite{hall1991freeway,chung2007relation}. 
On the other hand, microscopic traffic data such as vehicle trajectory data in a common reference system\cite{kovvali2007video, NGSIM}, helps researchers understand individual vehicle behaviors. 

Traffic data which captures observations of both microscopic and macroscopic phenomena has significant value in the transportation research community. A relative scarcity of these data has limited empirical validation of theory in individual dynamics and traffic system dynamics. To date, only a few datasets are available which capture both microscopic and macroscopic data. Data collected from mobile and fixed sensors by the Mobile Century experiments \cite{herrera2007traffic} have been utilized for traffic reconstruction. 
Additionally, INRIX and loop detector data \cite{kim2014comparing} have been collected and employed for cross-validation purposes. Purveyors of GPS-navigation applications have vehicle trajectory \cite{rempe2022estimation} and aggregate data, but these are proprietary. Further efforts and contributions are needed to expand the availability and diversity of datasets in this field. 

Ideally, microscopic and macroscopic scales in traffic measurement could be converted into another, similar to the concept of downsampling and upsampling in image processing \cite{acharya2005image}. In practice, transitioning from microscopic to macroscopic is straightforward through aggregation via Edie's definition \cite{edie1961}, whereas transforming from macroscopic to microscopic is more challenging due to the complex and heterogeneous dynamics of individual particles in the traffic flow system. The problem is often referred to as \textit{traffic reconstruction} \cite{herrera2007traffic, montanino2015trajectory, liu2021learning,thodi2022incorporating}. One paradigm from previous research involves calibrating sets of microscopic modeled driving behaviors to replicate macroscopic measurements or phenomena \cite{brockfeld2005calibration}. A limitation of using micro-simulation to reproduce macroscopic phenomena is the inherent mismatch between the model and real-world aggregated data, which may not fully capture the complex and dynamic nature of actual traffic systems \cite{punzo2014we, montanino2015trajectory}.
Another paradigm involves using kernel smoothing \cite{treiber2003adaptive,sanchez2023data} and similar techniques to interpolate and impute sparse observations, in order to enhance the resolution and completeness of traffic data by filling in gaps and smoothing out inconsistencies.  The adaptive smoothing method \cite{treiber2002reconstructing, treiber2003adaptive, treiber2011reconstructing} is a widely recognized approach in this direction. 
More recently, the refinement problem was proposed and defined by \cite{he2023refining}, which utilized a simple linear regression approach that accounts for different traffic states to enhance traffic measurement resolution. The method proposed in that study achieves a 2 by 2 lift in resolution for both space and time each time, demonstrating the potential to improve resolution from 200 meters by 60 seconds to 50 meters by 30 seconds. In other fluid dynamical systems, deep learning has been explored and applied to the reconstruction problem. These approaches have shown promise in accurately capturing and reconstructing complex fluid dynamics \cite{fukami2019super,yoda2020super,wang2022deep}, suggesting potential applicability to traffic flow systems \cite{morand2024deep} as well.

In recent years, the field of image processing has seen significant advancements due to generative artificial intelligence. Generative models have enabled the transformation of low-resolution images into high-resolution ones via a process known as \textit{super-resolution} \cite{park2003super}. This aligns well with the problem of enhancing traffic measurement resolution. The concept of generative deep learning gained prominence with the introduction of generative adversarial networks (GANs) \cite{goodfellow2020generative}, which have been extensively applied in traffic systems for tasks such as traffic data imputation \cite{chen2019traffic,zhang2021missing} and traffic state estimation \cite{mo2022trafficflowgan}.  More recently, diffusion probabilistic models have emerged as a promising approach \cite{ho2020denoising}. These models have been investigated for tasks such as noisy data recovery \cite{zheng2024recovering} and GPS trajectory data generation \cite{zhong2023guided}, demonstrating their robustness to hyperparameters and their potential for a wide range of applications. These advancements in generative diffusion models offer new opportunities for improving traffic measurement resolution and addressing the challenges associated with low spatio-temporal resolution data, which have not been investigated before.

\section{Methodology} \label{sec:method}

\subsection{Problem formulation}
In this article, stop-and-go wave refinement refers to the problem of recovering fine-grained speed data (high-fidelity) from corresponding coarse-grained traffic measurement data (relatively low-fidelity), which can be viewed as a super-resolution task. \textit{Coarse-grained data} typically collected from radar detectors or loop detectors, is considered low spatio-temporal resolution data and is denoted by $\mathbf{r} \in \mathcal{R}$. \textit{Fine-grained speed data}, generated from vehicle trajectory data, is viewed as high-fidelity data and is represented by $ \mathbf{m} \in \mathcal{M}$. For a given space and time range, forming tensors $\mathbf{r}$ and $\mathbf{m}$ herein, each observation of coarse-grained data is represented as $\mathbf{r} \in \mathbb{R}^{T_r \times S_r \times 3}$ and fine-grained data is represented as $\mathbf{m} \in \mathbb{R}^{T_m \times S_m \times 1}$, where $T_r,~S_r,~T_m,~S_m$ are the number of time and space observations in the space-time range, respectively. Typically, $T_r < T_m$ and $S_r < S_m$, which induces a dimension lifting in the output. The coarse-grained data contains three traffic measurements: speed, volume, and occupancy, whereas the fine-grained data includes only one measurement: speed. The problem addressed in this context is to find a mapping function \( f: \mathcal{R} \to \mathcal{M} \) that can restore the fine-scale details of stop-and-go waves in traffic from the coarse-grained data.
\subsection{Conditional denoising diffusion model}
The conditional denoising diffusion model is proposed by \cite{saharia2022image} to address the image super resolution tasks, and we apply this method here. Given a dataset consisting of input-output pairs of coarse-grained and fine-grained data, denoted as $\mathcal{D} = \{\mathcal{R}, \mathcal{M}\}$, which are samples from an unknown conditional distribution $p(\mathcal{M}\mid\mathcal{R})$, our objective is to learn a parametric approximation of $p(\mathcal{M}\mid\mathcal{R})$. This is achieved through a stochastic iterative refinement process that maps a coarse-grained traffic observation to a target fine-grained speed field.
\begin{figure}[H]
    \centering
    \includegraphics[width=\linewidth]{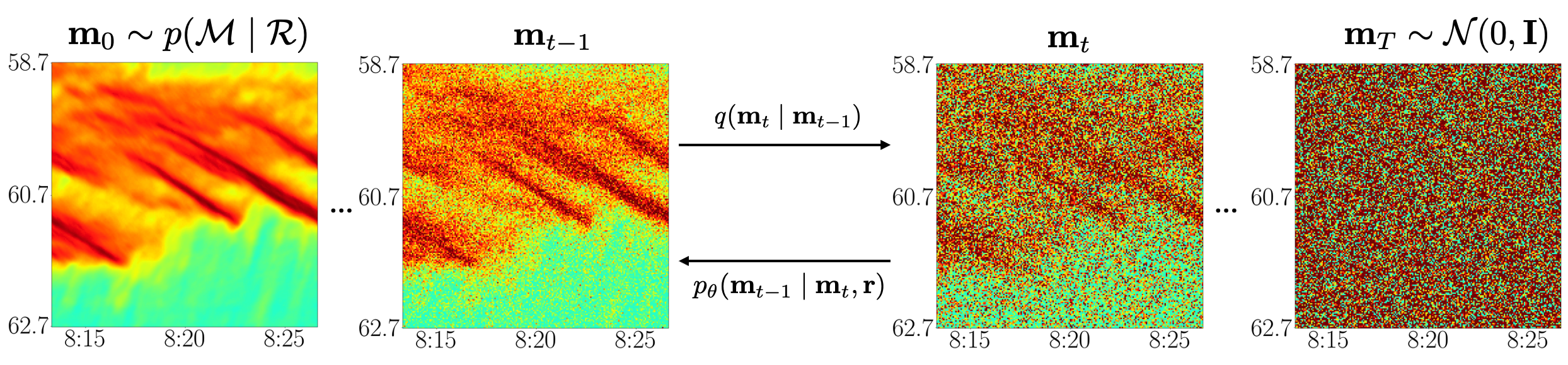}
    \caption{Demonstration of the forward (left to right) and inverse (right to left) diffusion process in our problem. Noted that the coarse-grained data is not shown here; refer to Figure \ref{fig:overview} for the input speed profile of $\mathbf{r}$.}
    \label{fig:model}
\end{figure}
We adapt the denoising diffusion probabilistic model (DDPM) \cite{saharia2022image} for conditional fine-grained speed field generation. The method contains a noising process (e.g. forward diffusion) and a denoising process (e.g. inverse diffusion) as shown in Figure \ref{fig:model}. These processes are defined as follows.
\vspace{0.35cm}

\noindent \textbf{Forward diffusion (add noise)}: the process $q$ gradually adds Gaussian noise over $T$ steps to the fine-grained data $\mathbf{m}_0 =\mathbf{m}$, which can be viewed as a forward Markovian diffusion process.
\begin{align}
q(\mathbf{m}_{1:T} \mid \mathbf{m}_0) &= \prod_{t=1}^{T} q(\mathbf{m}_t \mid \mathbf{m}_{t-1}), \\
q(\mathbf{m}_t \mid \mathbf{m}_{t-1}) &= \mathcal{N}(\mathbf{m}_t \mid \sqrt{\alpha_t} \mathbf{m}_{t-1}, (1 - \alpha_t) \mathbf{I}).
\end{align}
Here, the scalar parameters $\alpha_{1:T}$ are hyper-parameters where $0 < \alpha_t <1$,  which determine the variance of the noise added at each iteration $t$.
For each step, the distribution of $\mathbf{m}_t$ given by $\mathbf{m}_0$ can be expressed as:
\begin{equation}
    q(\mathbf{m}_t \mid \mathbf{m}_0) = \mathcal{N}(\mathbf{m}_t \mid \sqrt{\gamma_t} \mathbf{m}_0, (1 - \gamma_t) \mathbf{I}), \label{eq:noise}
\end{equation}
where $\gamma_t = \prod_{i=1}^{t}\alpha_i$.
\vspace{0.35cm}

\noindent \textbf{Learn the inverse diffusion (denoise)}: to reverse the forward diffusion process, a neural denoising model $f_\theta$ takes the coarse-grained data $\mathbf{r}$ and the noisy target data $\mathbf{\tilde m}$ as defined in~\eqref{eq:mtilde}:
\begin{equation}
    \mathbf{\tilde m} = \sqrt{\gamma} \mathbf{m}_0 + \sqrt{1-\gamma}\epsilon, ~~~\epsilon \in \mathcal{N}(0,\mathbf{I}) \label{eq:mtilde}.
\end{equation}
The model $f_\theta(\mathbf{r},~\mathbf{\tilde m},~\gamma)$ is trying to learn from the coarse data $\mathbf{r}$, the noisy target image $\mathbf{\tilde m}$, as well as the variance of the noise $\gamma$, to predict the noise vector $\epsilon$ in \eqref{eq:mtilde}. Here, $\gamma$ is the scalar parameter drawn from the distribution $p(\gamma)$. The objective function for training $f_\theta$ is written as:
\begin{equation}
    \mathbb{E}_{(\mathbf{r},\mathbf{m})} \mathbb{E}_{\epsilon, \gamma} \left\lVert f_{\theta}\left(\mathbf{r}, \underbrace{\sqrt{\gamma} \mathbf{m}_0 + \sqrt{1 - \gamma} \epsilon}_{\mathbf{\tilde{m}}}, \gamma \right) - \epsilon \right\rVert_{2}^{2} .
\end{equation}
During the training process, each step involves performing a gradient descent update to minimize the loss function and optimize the model parameters:
\begin{equation}
    \nabla_{\theta} \left\| f_{\theta}(\mathbf{r}, \sqrt{\gamma} \mathbf{m}_0 + \sqrt{1 - \gamma} \epsilon, \gamma) - \epsilon \right\|_2^2,
\end{equation}
 until convergence. 
The detailed proof and justification of the choice of the objective function for $f_\theta$ from the perspective of a variational lower bound and denoising score-matching can be found in \cite{ho2020denoising, saharia2022image}. 

Note that the model is learning the output of $f_\theta$ to the Gaussian noise, not the $\mathbf{m_0}$. As a result, it is not feasible to use the typical tricks to get physics informed learning \cite{cai2021physics}. To address this, we design a weighting function $\mathbf{W}(\mathbf{m},v_c)$ for the loss to focus on the areas of interest, which, in our problem, are the stop-and-go waves. The weighting function is written as:
\begin{equation}
\mathbf{W}_{ij}(\mathbf{m},v_c) = \begin{cases} 
\omega & \text{if } \mathbf{m}_{ij} < v_c \\
1-\omega & \text{if } \mathbf{m}_{ij} \geq v_c 
\end{cases},
\end{equation}
where $\omega$ and $v_c$ is the hyperparameter, $\mathbf{m}_{ij}$ denotes the value at time index $i$ and space index $j$ in the fine-grained speed data.
The weighting matrix is subsequently integrated into the training process. The weighted gradient descent step during training is defined as:
\begin{equation}
    \nabla_{\theta} \left\| \mathbf{W} \odot (f_{\theta}(\mathbf{r}, \sqrt{\gamma} \mathbf{m}_0 + \sqrt{1 - \gamma} \epsilon, \gamma) - \epsilon) \right\|_2^2 , 
\end{equation}
where \( \odot \) denotes element-wise multiplication.
\vspace{0.35cm}

\noindent \textbf{Inference via iterative refinement}: The reverse inference process $p$ iteratively denoises the noisy data using conditioned on the input coarse-grained data $\mathbf{r}$. Starting with purely Gaussian noise $\mathbf{m}_T \sim \mathcal{N}(0, \mathbf{I})$ and the corresponding coarse-grained data $\mathbf{r}$, the model iteratively refines the speed field through successive iterations $(\mathbf{m}_{T-1}, \cdots,  \mathbf{m}_{t}, \mathbf{m}_{t-1}, \cdots, \mathbf{m}_{0})$ according to the learned conditional transition distribution:
\begin{equation}
    \mathbf{m}_{t-1} = \frac{1}{\sqrt{\alpha_t}} \left( \mathbf{m}_t - \frac{1 - \alpha_t}{\sqrt{1 - \gamma_t}} f_{\theta}(\mathbf{r}, \mathbf{m}_t, \gamma_t) \right) + \sqrt{1 - \alpha_t} z ,
    \label{eq:iter}
\end{equation}
where $z$ is defined as Gaussian noise for all the steps other than $t=1$ (i.e.,  $z \sim \mathcal{N}(0, \mathbf{I}) \text{ if } t > 1, \text{ else } z = 0$). The derivation of~\eqref{eq:iter} can be found in \cite{saharia2022image}.

\vspace{0.35cm}

\noindent \textbf{Neural denosing model choice}: the UNet \cite{ronneberger2015u} architecture is a popular choice for diffusion denoising model regression \cite{ho2020denoising,saharia2022image} because its encoder-decoder structure with skip connections effectively captures multi-scale features and preserves high-resolution details throughout the denoising process. Detailed task specific architectural details can be found in Section~\ref{sec:exp}.

\section{Data} \label{sec:data}
To address the aforementioned problem, a sufficient amount of high-fidelity data is essential, particularly for training a deep generative model. The data used in this paper is collected from two large-scale traffic measurement systems: a network of millimeter-wave \textit{radar detection systems} (RDS) spanning 17.1 miles at roughly 0.3 to 0.5 mile spacing, and a camera network capable of capturing massive vehicle trajectory data densely covering 4.2 miles of roadway. Each system is detailed below.

\subsection{I-24 MOTION data}
I-24 MOTION \cite{gloudemans202324} is a traffic instrument located along 4.2 miles of interstate roadway near Nashville, Tennessee. It consists of 276 4k-resolution cameras mounted on 40 110-foot tall traffic poles, sufficient to densely survey this portion of roadway. Computer vision detection and tracking algorithms \cite{gloudemans2024so} are applied to the video data to produce vehicle trajectories for vehicles travelling through the roadway.

\begin{figure}[H]
    \includegraphics[width = \textwidth]{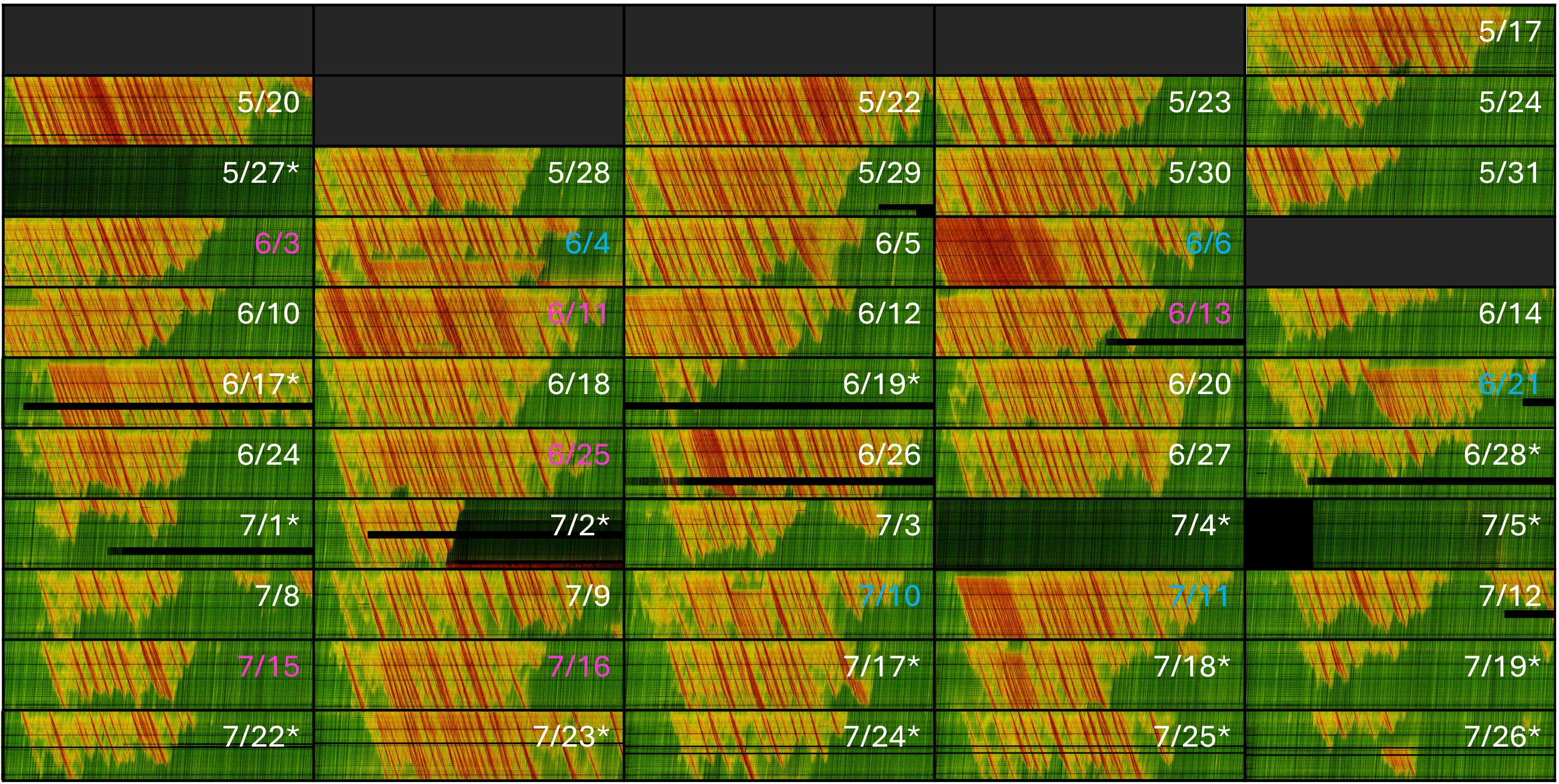}
    \caption{Time-space trajectory data diagrams for each included day in the I-24 MOTION: \texttt{WaveX} dataset. Each day of data spans 4.2 miles (y-axis) and 4 hours (x-axis) and is produced with $\sim$ 2-foot positional accuracy and at 10Hz. Days with no data due to e.g. system maintenance are filled with dark grey. (pink text) validation holdout days. (blue text) testing holdout (notable event) days. Days with an asterisk were not used in model training and evaluation due to missing data or other anomalous events, but are released with the dataset.}
    \label{fig:MOTION-raw}
\end{figure}

Trajectory data produced by I-24 MOTION was used to create fine-grained \textit{mean speed field} data for this work. A selection of 33 days were selected, recorded from May 17 to July 17, 2024; for each day the morning rush hour period (6:00 or 5:30 AM to 10:00 AM) was considered as this period has the most dynamic and variable traffic conditions.  Figure \ref{fig:MOTION-raw} shows the raw trajectory data for each selected day. The selected trajectory data was then sampled to create a fine-grained mean speed field with 4 second and 0.02 mile bin resolution. For cells with no data and cells with inconsistent values, speed was interpolated using an adaptive smoothing method \cite{treiber2003adaptive,ji2024virtual,ji2024scalable}, with parameters shown in Table \ref{tab:asm}. 

\begin{table}[H]
\centering
\caption{Summary of I-24 MOTION data statistics}
\vspace{0.2cm}
\begin{tabular}{rrrrrrr}
\toprule
      & \textbf{VMT} (mile)       & \textbf{VHT} (hour)      & \textbf{MS} (mph)  & \textbf{Training}   & \textbf{Validation}   & \textbf{Testing}   \\ \hline
count &         &       &      & 21     & 7      & 5      \\
mean  & 69270 & 2166 & 33.2 & 33.9 & 34.5 & 28.6 \\
st.dev.   & 5486  & 350  & 8.0  & 7.7  & 9.3  & 7.2  \\
min   & 56158 & 1456 & 19.8 & 21.1 & 22.6 & 19.8 \\
25\%  & 65445 & 1941 & 28.5 & 29.9 & 29.5 & 24.3 \\
50\%  & 70682 & 2198 & 31.5 & 31.5 & 32.6 & 27.6 \\
75\%  & 72538 & 2326 & 38.0 & 38.0 & 38.4 & 33.2 \\
max   & 78825 & 2856 & 52.4 & 52.4 & 50.4 & 38.1 \\ \bottomrule
\end{tabular}
\label{tab:stats} 
\end{table} 
Table \ref{tab:stats} provides an overview of the key metrics in the I-24 MOTION dataset, including Vehicle Miles Traveled (VMT), Vehicle Hours Traveled (VHT), and the mean speed (MS, the vehicle-miles-weighted speed) \cite{chen2001freeway}. It also outlines the strategy for splitting the data into training and validation sets, with a 75\% allocation for training and 25\% for validation. 
Note that the testing set comprises data from 5 days with distinctive congestion patterns due to observable events, which are excluded from both the training and validation sets to avoid model training imbalance. The validation set  is used during training to tune hyperparameters and monitor the performance, providing feedback to prevent overfitting (e.g. early stopping). The test set is kept completely separate from the training process and is used only to evaluate the model generalization on unseen data.

\begin{figure}[H]
    \centering
    \includegraphics[width=\linewidth]{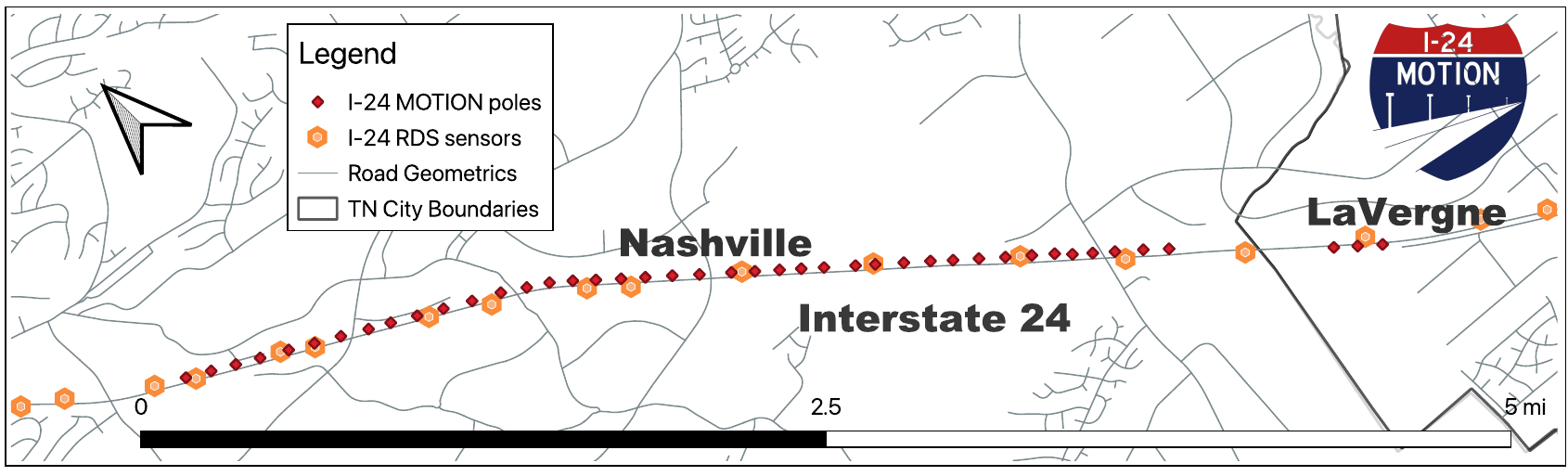}
    \caption{Locations of I-24 MOTION camera poles (which provide complete roadway coverage between) and I-24 RDS sensor placement.
    }
    \label{fig:rds}
\end{figure}

\subsection{I-24 RDS data}
The I-24 RDS data is obtained from the I-24 radar detection system (RDS) \cite{zhang2023marvel,coursey2024ft}. The system collects lane-level local traffic measurements including speed, occupancy, and volume, aggregated at 30-second intervals. All the radar sensors utilized in this study were meticulously calibrated. For the purposes of this study, we specifically selected the subset of sensors that align spatially with the I-24 MOTION testbed as shown in Figure \ref{fig:rds}. A total of 11 RDS sensors are aligned in this section. Two additional sensors located at both ends are selected to help with reconstruction.
\subsection{Dataset processing}
\noindent\textbf{Pre-processing.} To address the sparsity of I-24 RDS data at higher space-time resolutions, we utilize the adaptive smoothing method \cite{treiber2003adaptive}. This widely-used upsampling strategy for traffic measurements lifts the dimensionality of the RDS data to match that of I-24 MOTION, specifically to $\mathbb{R}^{T_m \times S_m \times 3}$, denoted as $\mathbf{\tilde r}$. The adaptive smoothing method produces a complete matrix, which is then used as input data for training and validation. Parameters used in the adaptive smoothing method is listed in Table \ref{tab:asm} In this paper, both $S_m$ and $T_m$ are set to 200, corresponding to a spatio-temporal tensor spanning 800 seconds and 4 miles. In this paper, experiments and discussions are exclusively based on data from lane 1 (the leftmost lane) and the high-occupancy vehicle (HOV) lane for both the I-24 MOTION and I-24 RDS data. 

\begin{table}[H]
\centering
\caption{Parameters of the adaptive smoothing method}
\label{tab:asm}
\begin{tabular}{llcc}
\toprule
         & Meaning & I-24 MOTION & I-24 RDS \\ \hline
$\sigma$(mile) &  smoothing width in time coordinate  & \textcolor{red}{0.02} &    \textcolor{red}{0.10}     \\
$\tau$(second) & smoothing width in space coordinate           & \textcolor{red}{4} &   \textcolor{red}{9}      \\ 
$c_{\text{free}} $(mph) & wave speed in free traffic        &\textcolor{red}{-12}  & \textcolor{red}{-12}       \\ 
$c_{\text{cong}} $(mph)& wave speed in congested traffic        &\textcolor{red}{45}&\textcolor{red}{45}         \\ 
$V_{\text{thr}}$(mph) & crossover from congested to free traffic        &\textcolor{red}{40} &\textcolor{red}{40}    \\ 
$\Delta V$(mph) & transition width between congested and free traffic        &\textcolor{red}{10}  & \textcolor{red}{10}    \\ 
\bottomrule
\end{tabular}
\end{table}

\noindent \textbf{Dataset augmentation.} To augment the dataset, we employed a sliding window strategy, generating multiple overlapping sub-datasets from the original daily dataset matrix. Starting with an original tensor of size 3600 $\times$ 200, we created multiple 200 $\times$ 200 sub-tensors. A 200 $\times$ 200 window slides across the larger matrix with a defined step size, producing a new 200 $\times$ 200 tensor at each step. In this study, we used a sliding window step size of 10, corresponding to 40 seconds in time. For the evaluation part, the augmentation via sliding window is not used.

\subsection{Data Availability}
The dataset associated with this paper, \texttt{WaveX}, would be a new release compared to our previous INCEPTION dataset. All days of high-resolution mean speed field data, as well as all corresponding days of RDS data, will be made available at \url{i24motion.org}.
\section{Experiments}
\label{sec:exp}
\subsection{Task specific architectural details}
\begin{figure}[H]
    \centering
    \includegraphics[width=\linewidth]{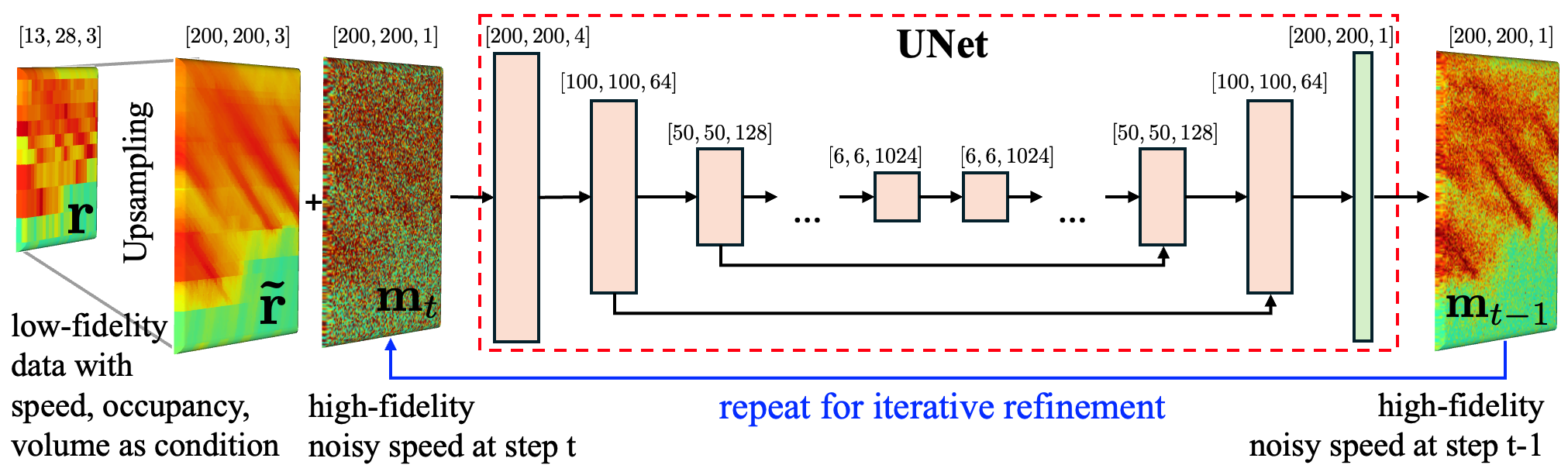}
    \caption{The UNet architecture iteratively refines the denoising process by concatenating the output from the previous step $\mathbf{m}_{t}$ with the upsampled low spatio-temporal resolution data $\mathbf{\tilde r}$. This process is conditioned on speed, occupancy, and volume of the low spatio-temporal resolution data. The specific UNet architecture employed for our tasks is depicted in the middle of the figure.}
\label{fig:UNet}
\end{figure}

In our problem, the coarse-grained data $\mathbf{r}$ is preprocessed with upsampling to be $\mathbf{\tilde r}$ (as shown in Figure \ref{fig:UNet}). The upsampled data is then concatenated with the noisy fine-grained speed data $\mathbf{m}_t$. This process generates a 4-channel tensor as the model input, while the model output is a single-channel tensor $\mathbf{m}_{t-1}$.
The UNet \cite{ronneberger2015u} architecture employed in this paper comprises an initial level with 64 channels, followed by levels with 128, 256, and 512 channels, culminating in a bottleneck level with 1024 channels, as shown in Figure \ref{fig:UNet}. Each block within this architecture consists of three convolutional ResNet blocks augmented with a multi-head attention mechanism, leading to a total of 130 million trainable parameters.

In this paper, we define a sequence of $\gamma$ values, which are essential for our diffusion process. These values are uniformly distributed with $p(\gamma)$ within a specified range for $T$ steps, starting from $\gamma_{\text{start}} = 0.001$ to $\gamma_{\text{end}} = 0.02$. The total number of values in this sequence is equal to the number of time steps, $T=500$, in our diffusion process. The learning rate for the optimizer of the neural denoising model is set at $2\times10^{-4}$. The velocity $v_c$ is set to 15 mph, and the parameter $\omega$ is set to 0.8. An early stopping strategy is employed when the validation loss begins to increase. The best performance on the validation dataset is retained for inference purposes.

\subsection{Benchmarked methods}
For the purpose of comparison, several methods evaluated on the formulated problem:
 \begin{figure}[H]
    \centering
    \includegraphics[width=0.8\linewidth]{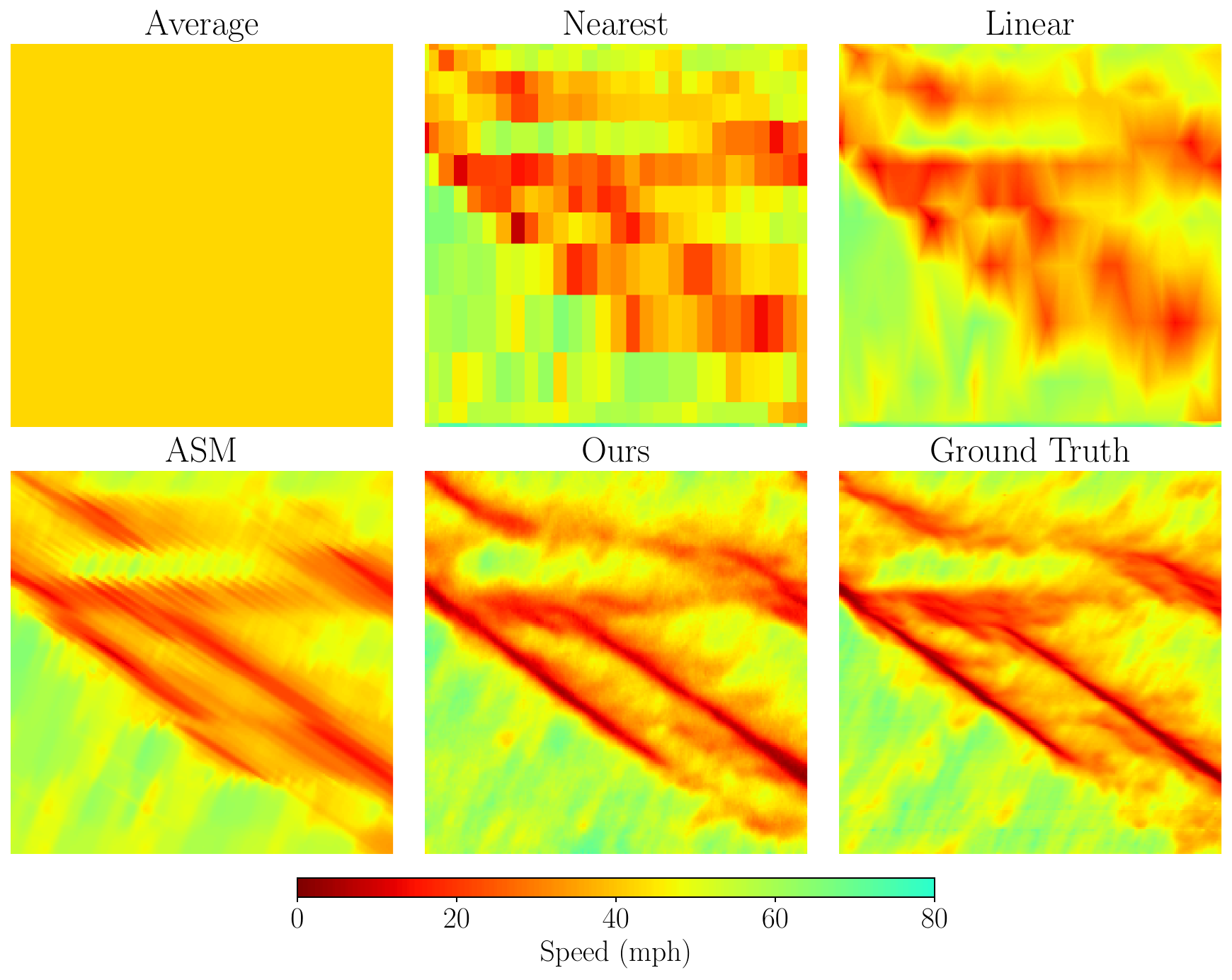}
    \caption{Comparison of baseline methods and our reconstruction against the ground truth.}
    \label{fig:demo-baseline}
\end{figure}
\begin{enumerate}[label=(\roman*),noitemsep]
\item \textbf{Average:} All data points are computed as the average of the available samples, under the assumption that the data is uniformly distributed across both space and time.
\item \textbf{Nearest:} fills in missing data points by assigning the value of the nearest available data point.
\item \textbf{Linear:} linear interpolation of missing data points (interpolation is carried out across the time dimension first, then the space dimension).
 \item \textbf{Adaptive smoothing method (ASM) \cite{treiber2003adaptive,treiber2011reconstructing}:} interpolates and smooths the values using two different smoothing kernels: one for free-flowing traffic and the other for congested traffic. The results are then combined using a $\tanh$ activation function.
 \item \textbf{Ours:} the generative model with iterative refinement proposed in this paper.
\end{enumerate}

A visual demonstration of each baseline method, alongside our reconstruction and the ground truth, is presented in Figure~\ref{fig:demo-baseline}.

\subsection{Code and tools}
The model is coded using PyTorch \cite{paszke2017automatic}, leveraging CUDA for accelerated computation. All experiments are performed on a machine with two NVIDIA RTX A6000 GPUs and AMD Ryzen Threadripper 3960X 24-Core Processor. 
The code will be open-sourced
, facilitating future research and enabling other researchers to build upon our work.

\begin{table}[H]
\centering
\caption{Evaluation matrices on baselines highlighting our improvements over the second-best approach. In the results, the second-best method is indicated with \underline{underline}, while the best-performing method is marked in \textbf{bold}.}
\vspace{0.2cm}
\begin{tabular}{lcccccc}
\toprule
Metric & Average & Nearest & Linear & ASM & Ours & Improvement  \\ 
\midrule
\multicolumn{6}{l}{\textbf{Training}} \\
\hline
WD & 20.88   & \underline{1.90}    & 2.72   & 2.36 &\textbf{1.05} & 47.22\%\\
RMSE  & 23.40   & 9.43    & 8.01   & 6.89 &\textbf{4.44} & 35.49\% \\
MAPE  & 1.50    & 0.48    & 0.49   & 0.37 &\textbf{0.14} & 62.57\%\\
\midrule
\multicolumn{6}{l}{\textbf{Validation}} \\
\hline
WD & 20.51   & \underline{1.94}    & 2.75   & 2.36 & \textbf{1.28}   &34.02\% \\
RMSE  & 23.00   & 9.48    & 8.09   & \underline{6.97} & \textbf{5.94} & 14.81\% \\
MAPE  & 1.45    & 0.48    & 0.48   & \underline{0.37} & \textbf{0.25} & 32.18\% \\
\midrule
\multicolumn{6}{l}{\textbf{Testing}} \\
\hline
WD & 19.68   & \underline{1.79}    & 2.59   & 2.27 & \textbf{1.48} & 17.32\% \\
RMSE  & 22.33   & 9.82    & 8.35   & \underline{7.22} & \textbf{6.25}& 13.40\%\\
MAPE  & 1.55    & 0.53    & 0.54   & \underline{0.43} & \textbf{0.28}& 34.87\%\\
\bottomrule
\end{tabular}
\label{tab:results}
\end{table}

\subsection{Evaluation Metrics}
The following  metrics were used to evaluate the benchmarked methods:

\begin{enumerate}[label=(\roman*),noitemsep]
\item \textit{Wasserstein distance} (WD) \cite{vallender1974calculation}: The Wasserstein distance quantifies the difference between two probability distributions. A smaller distance indicates a closer alignment with the reference distribution.  
\item \textit{Root Mean Squared Error} (RMSE): Unit in miles per hour (mph), RMSE measures the average magnitude of the errors between reconstructed and actual values, with greater weight assigned to larger errors. Lower RMSE values indicate more accurate reconstructions.  
\item \textit{Mean Absolute Percentage Error} (MAPE): Reported as a percentage, MAPE reflects the average absolute difference between reconstructed and actual speed relative to the actual speed. Lower MAPE values indicate more accurate reconstructions.  
\end{enumerate}

\section{Results} \label{sec:result}
\subsection{Overall performance}
Table \ref{tab:results} presents a detailed evaluation comparing the performance of our method with various baselines. The results clearly indicate that our method outperforms the baseline methods across all evaluation metrics and datasets. Notably, the superior performance of our method in minimizing the \textit{Wasserstein Distance} (WD), 47.22\% in training, 34.02\% in validation and 17.32\% in testing over the second-best methods, underscores its efficacy in accurately reconstructing speed distributions, which is crucial for stop-and-go wave reconstruction. The substantial improvement in the WD metric suggests that our method captures the correct patterns, even if there are minor spatial misplacement, which is an acceptable limitation as confirmed by the travel time estimates.

In terms of RMSE, our method demonstrates the improvements, with a 35.49\% advantage in training, 14.81\% in validation, and 13.40\% in testing, indicating its robustness in producing refined data with minimal error. The MAPE metric reveals even more pronounced benefits, with our method achieving a 62.57\% advantage in training, 32.18\% in validation, and 34.87\% in testing. These results highlight the overall efficacy of our approach in improving traffic reconstruction across various metrics.

\begin{figure}[H]
    \centering
    \includegraphics[width=\linewidth]{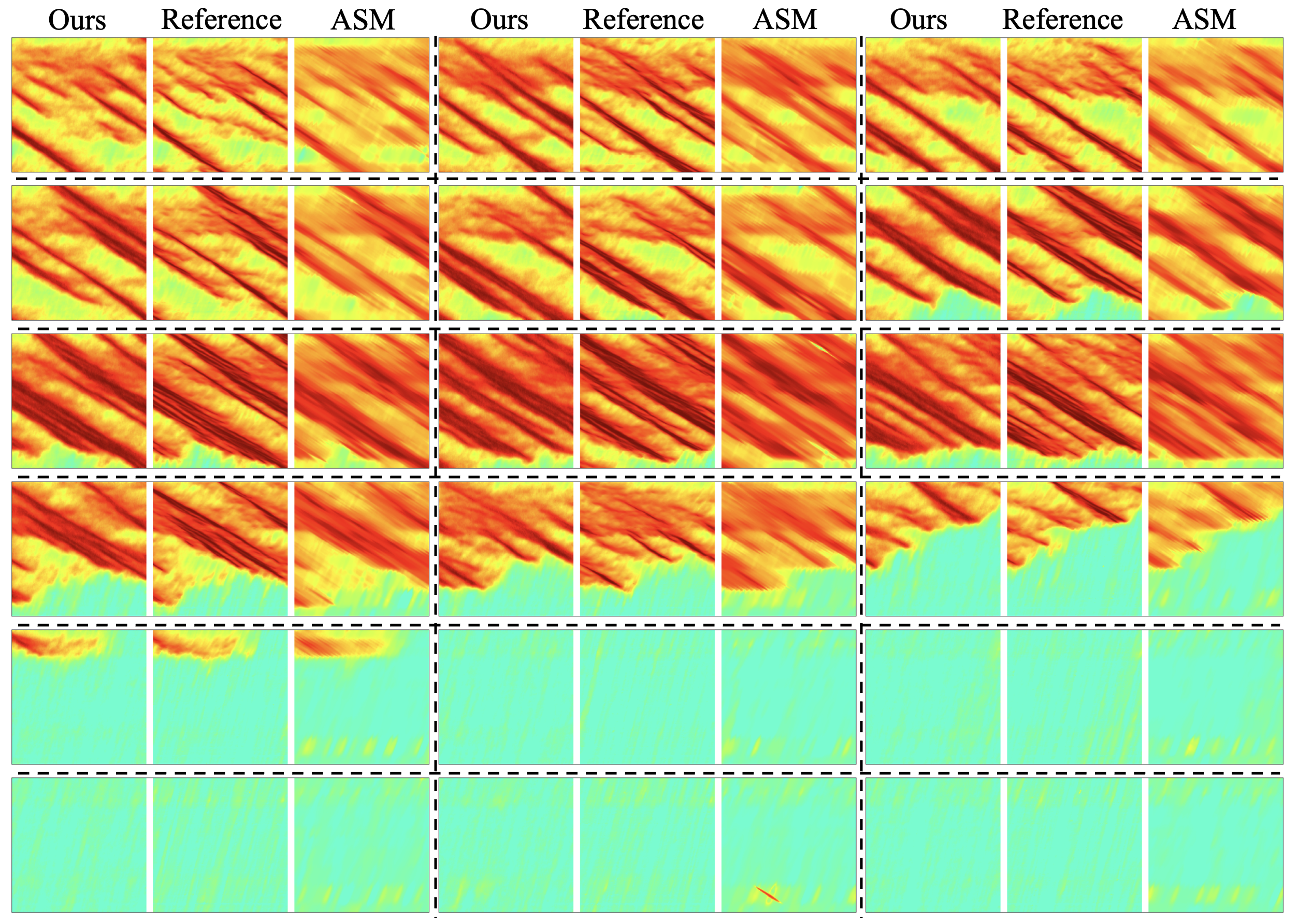}
    \caption{Stop-and-go waves reconstruction on validation data, (June 3, 2024). Each image contains data from 800 seconds (x-axis) and 4 miles (y-axis).}
    \label{fig:val}
\end{figure}
\subsection{Congested traffic reconstruction}
\begin{figure}[H]
    \centering
    \includegraphics[width=\linewidth]{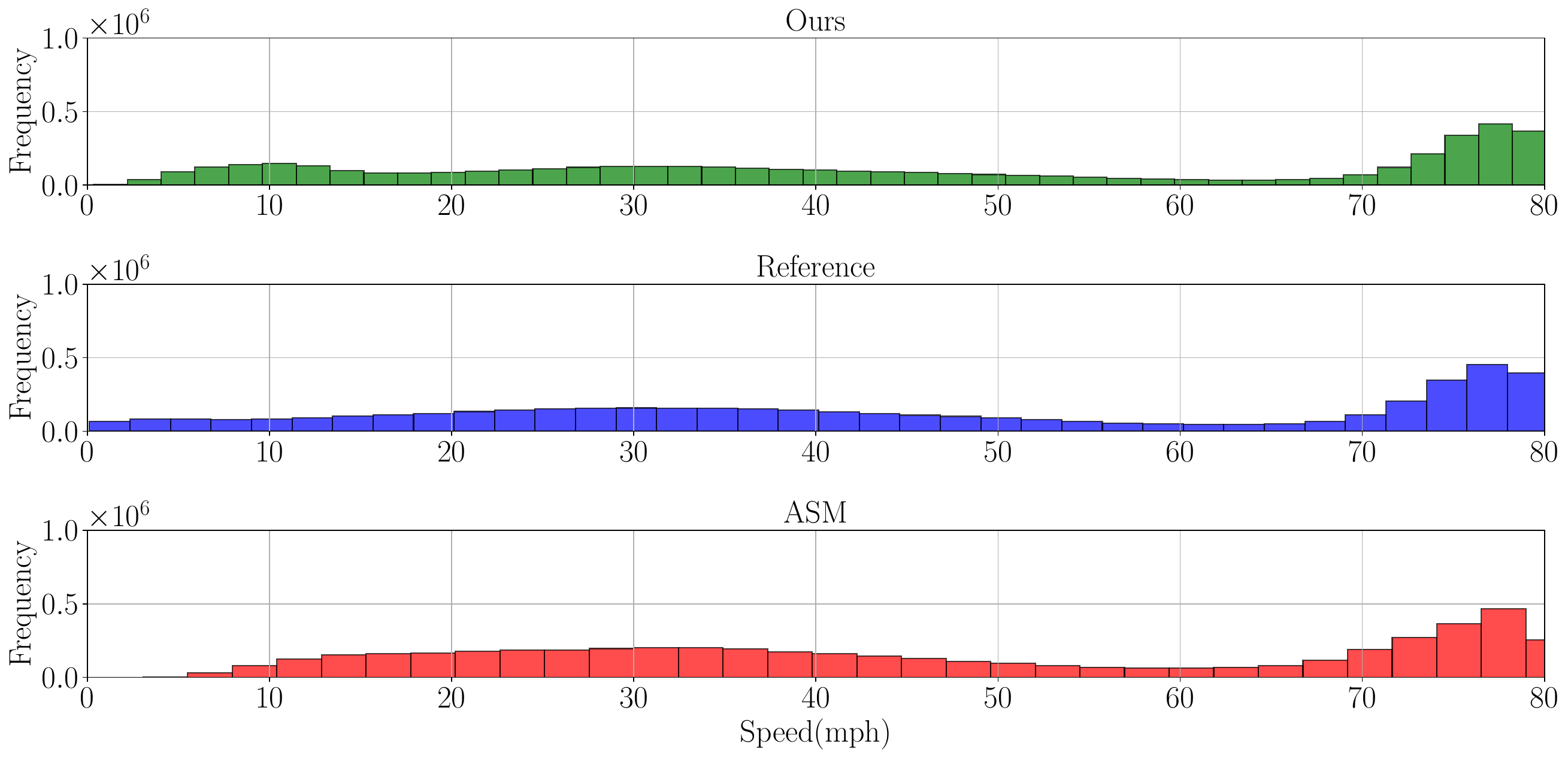}
    \caption{Distribution of accumulated speed profiles on the validation dataset: ours (top), reference (middle), ASM (bottom)}
    \label{fig:freq_val}
\end{figure}

Figure \ref{fig:freq_val} displays the distribution of the speed profile for the reconstruction from ours (top) and ASM (bottom), with the reference in the middle, on the validation dataset. Note that values outside the 0-80 mph range are clipped to this range. As can be observed from the ASM reconstruction speed profile distribution, the RDS sensor data with ASM fails to capture the low-speed range, particularly in the 0-10 mph range. 

In contrast, our method demonstrates promising reconstruction performance in the low-speed traffic segments, which correspond to stop-and-go traffic conditions. The improved performance in the low-speed range suggests that ours is more effective in reconstructing the intricate patterns of stop-and-go waves, enhancing the overall fidelity of the traffic measurements.
\subsection{Distribution of errors}
\begin{figure}[H]
    \centering
    \includegraphics[width=0.9\linewidth]{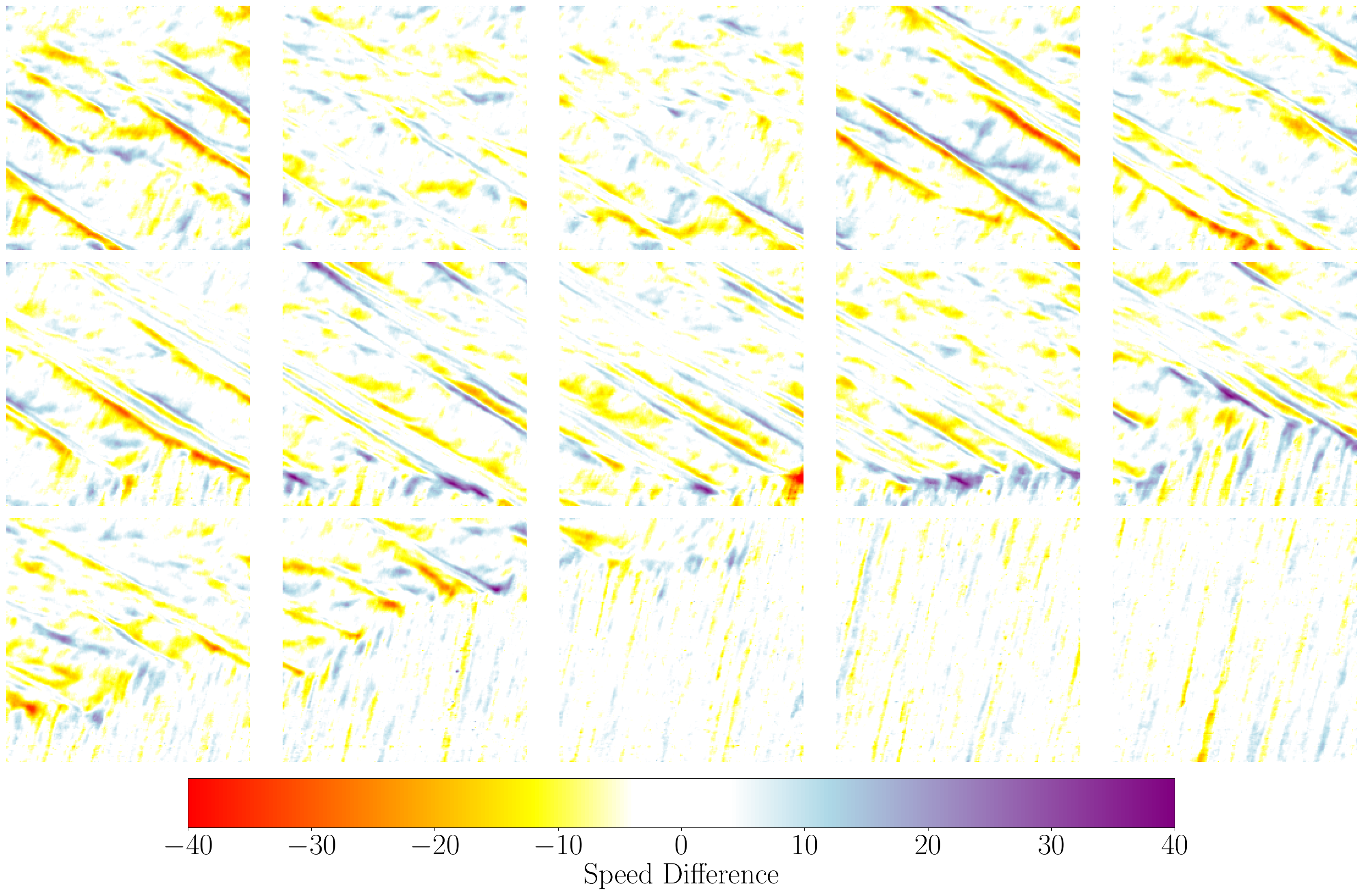}
    \caption{Error distribution plot for the reconstructed speed and the reference speed field, dated on the validation dataset, June 3, 2024
    }
    \label{fig:error}
\end{figure}

Despite the better results compared to baselines of our method indicated by the RMSE and MAPE in Table \ref{tab:results}, the errors do not align as closely with the distribution of the speed profile as anticipated. To investigate this discrepancy, we plot the absolute error (shown in Figure \ref{fig:error}) on the space-time diagram to analyze the spatio-temporal distribution of our error terms and gain deeper insights into the underlying causes.

As can be seen in Figure \ref{fig:error}, the error terms in speed difference generally exhibit a diagonal pattern, alternating in a faster and slower fashion. This may indicate that the spatio-temporal locations of the stop-and-go waves are mismatched. Specifically, this pattern suggests that the reconstructed data may not accurately align with the true positions and time of the stop-and-go waves. Such mismatches could be due to various factors, including the limitations of the low spatio-temporal resolution data in capturing rapid fluctuations in traffic speed or the inherent challenges in the modeling process. However, we notice that this does not compromise the detection of the ``patterns" as confirmed by the high performance with respect to the Wasserstein distance (Table \ref{tab:results}).

\subsection{Travel time reconstruction}
By utilizing the generated speed field, travel time can be estimated through the use of virtual vehicles \cite{tsanakas2022generating, ji2024virtual}. For each scenario, a number of vehicles are introduced into the speed field to create virtual trajectories, allowing for the estimation of travel time through these virtual vehicles. In our case, virtual vehicles are deployed into the speed field every 10 seconds, with the corresponding virtual trajectories updated at a frequency of 1Hz. 

\begin{figure}[H]
    \centering
    \includegraphics[width=\linewidth]{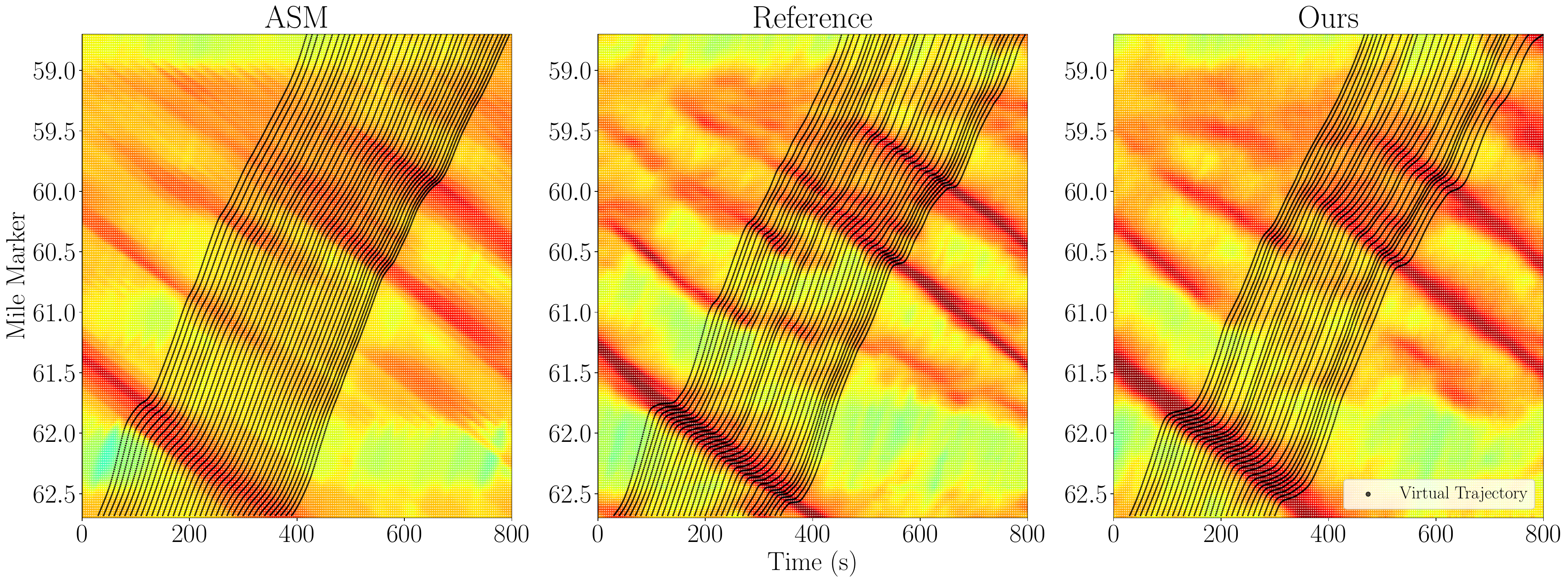}
    \caption{Travel time estimation via virtual trajectories}
    \label{fig:vt}
\end{figure}

For the fair comparison, only complete trajectories, in both space and time (as shown in Figure \ref{fig:vt}), are retained for analysis. This means that only those trajectories with an average speed exceeding 18 mph are considered, given that our generated field spans 4 miles and 800 seconds (the slope of the diagonal is 18 mph). For each speed field, the travel time is estimated by averaging the travel times of all complete virtual trajectories.

\begin{table}[H]
\centering
\caption{Statistics on the travel time (in seconds) estimation on the validation dataset}
\label{tab:tt}
\vspace{0.2cm}
\begin{tabular}{lccc}
\toprule
                              & Reference & ASM            & Ours          \\ \hline
\textbf{18-30 mph} (36 total)  &           &                &               \\ \hline
mean (seconds)                 & 579.9     & 512.1         & 596.7         \\
relative error                & -         & \textbf{-12\%} & \textbf{+3\%} \\ 
\midrule
\textbf{30-50 mph} (25 total) &           &                &               \\ \hline
mean (seconds)              & 385.0     & 360.5          & 390.5         \\
relative error                & -         & \textbf{-6\%}  & \textbf{+1\%} \\
\midrule
\textbf{50+ mph} (43 total)   &           &                &               \\ \hline
mean (seconds)             & 198.2     & 200.9          & 197.3         \\
relative error                & -         & \textbf{+1\%}  & \textbf{+0\%} \\ 
\bottomrule
\end{tabular}
\end{table}
The travel time estimation results across the validation dataset are summarized in Table \ref{tab:tt}, categorized by the average speed of virtual vehicles into three groups: 18-30 mph, 30-50 mph, and greater than 50 mph. This categorization provides a comprehensive analysis of the performance of our model in different speed ranges.

For the 18-30 mph speed range, which includes a total of 36 samples, the mean travel time estimated by our model is 596.7 seconds with a standard deviation of 81.3 seconds. Compared to the reference mean travel time of 579.9 seconds, our model shows a relative error of +3\%, whereas the ASM model exhibits a relative error of -12\%. This indicates that while our model slightly overestimates the travel time in this speed range, it does so with reasonable accuracy. In the 30-50 mph speed range, consisting of 25 samples, our model estimates a mean travel time of 390.5 seconds with a standard deviation of 63.8 seconds. The reference mean travel time is 385.0 seconds, resulting in a relative error of +1\% for our model. The ASM model, on the other hand, has a relative error of -6\%. This suggests that our model performs with high accuracy in this speed range, with minimal deviation from the reference values. The findings suggest that our model is robust across various speed ranges, especially good in dealing with the challenging stop-and-go traffic conditions.
\subsection{Wave speed estimation}
We apply the wave analysis tools developed in \cite{ji2024enabling,ji2024scalable} to validate the wave speed for wave front and tails. Wave fronts are identified as the spatio-temporal points where a virtual vehicle speed drops to 15 mph, while wave tails are defined as the points where the speed increases back above 15 mph.

\begin{table}[H]
\centering
\caption{Wave fronts and tails estimated from the speed field}
\label{tab:wave_speed}
\begin{tabular}{lccc}
\hline
              & Reference & ASM   & Ours  \\ \hline
\textbf{wave front}    &           &       &       \\ \hline
mean speed (mph)    & 12.01     & 12.22 & 12.69 \\
speed st.dev. (mph) & 2.14      & 1.82  & 2.49  \\ \hline
\textbf{wave tail}     &           &       &       \\ \hline
mean speed (mph)   & 11.24     & 12.16 & 11.81 \\
speed st.dev. (mph) & 1.65      & 1.90  & 1.88  \\ \hline
\end{tabular}
\end{table}

As can be seen from the Table \ref{tab:wave_speed}, our reconstructed wave speed is very close to the ASM, which is predefined as 12 mph when preprocessing the low spatio-temporal resolution data. Our reconstruction captures the varying speeds of the wave fronts and wave tails, highlighting its effectiveness in reconstructing the evolving patterns of stop-and-go waves. In contrast, the ASM produces relatively consistent speeds for both the wave front and tail, reflecting its predefined assumptions rather than the dynamic nature of real traffic patterns. However, there is a mismatch in the estimated mean speed between the wave fronts and wave tails. Addressing this mismatch may require further development, potentially by incorporating mechanisms to explicitly enforce wave speed consistency, such as integrating relevant constraints into the loss function.

\section{Discussions}
\label{sec:discussion}
\subsection{Events and crashes}
\begin{figure}[H]
    \centering
    \includegraphics[width=0.5\linewidth]{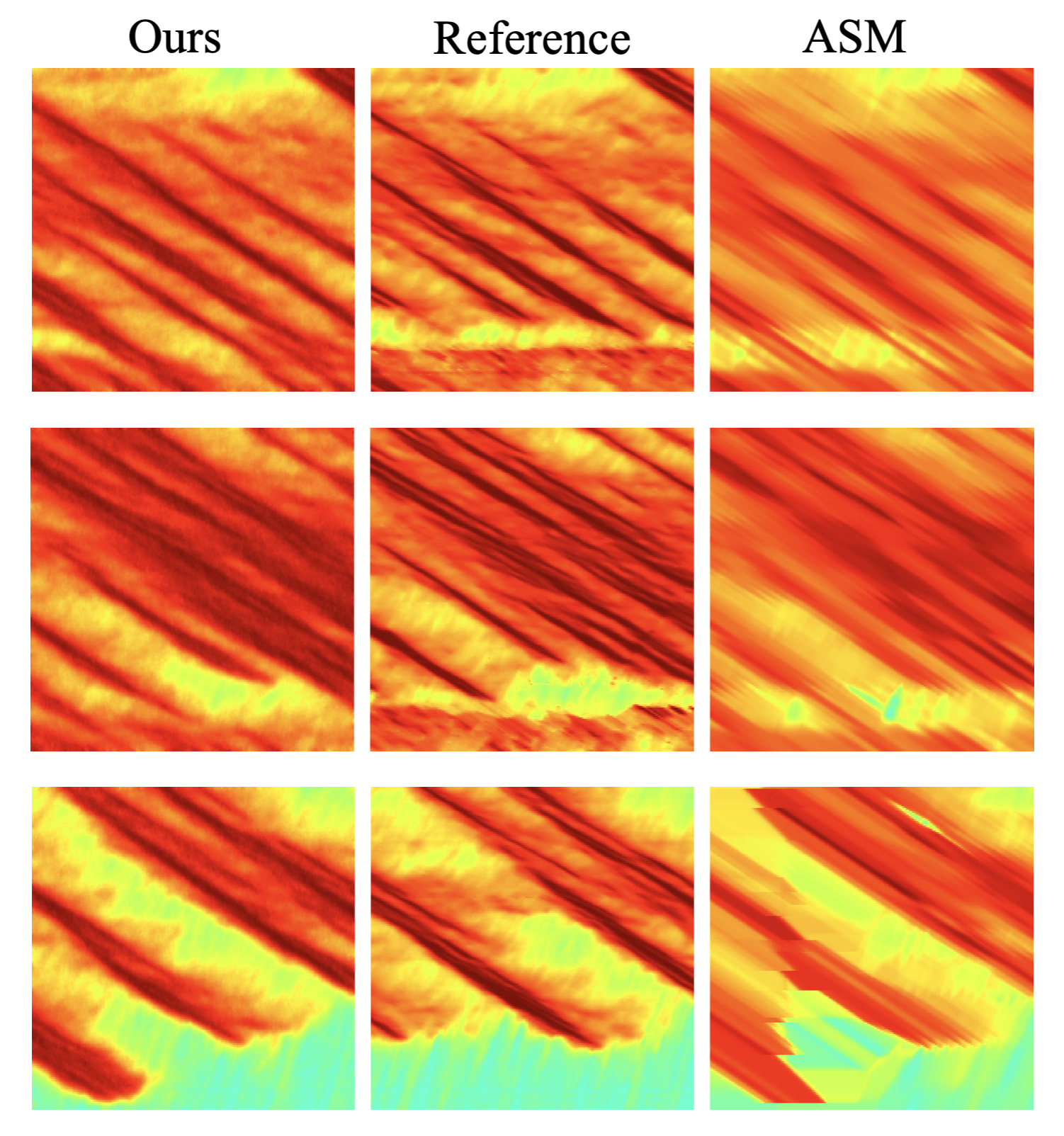}
    \caption{Failed examples in the validation dataset: crashes not observed by the conventional sensor (top and middle), and missing conventional sensor data (bottom).}
    \label{fig:fail_val}
\end{figure}

We conducted a manual inspection of all outputs from the validation dataset to identify any apparent errors produced by our models. Figure \ref{fig:fail_val} shows 3 of the typical failed examples in the validation datasets. The initial two rows of the examples indicate that there are crashes/events concealed within the stop-and-go waves. These are not detectable by conventional sensors and, consequently, cannot be retrieved using our methods, which is acceptable. Future research could explore the potential of incorporating event context information into the model as a form of prompt for generation.
The final row of the example indicates that data is missing for half of the observation period. The ASM interpolation leverages the non-compact kernel to extend the reconstructed waves; however, this leads to an overestimation of wave propagation, which in turn causes our method to reconstruct the wave in the leftmost section for a longer duration than observed.

\begin{figure}[H]
    \centering
    \includegraphics[width=0.75\linewidth]{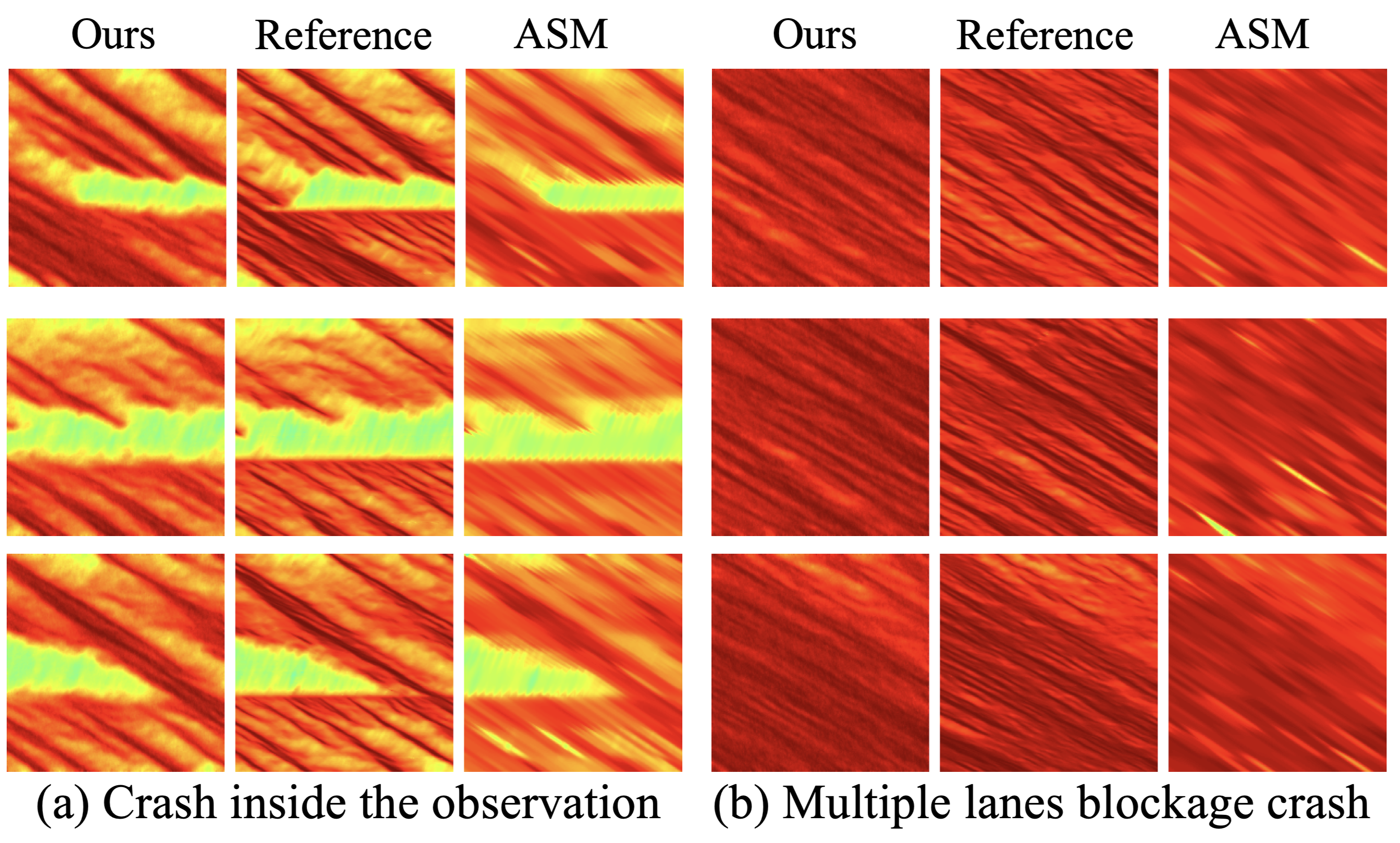}
    \caption{More examples for crash scenarios in the test dataset: (a) crashes inside the stop-and-go traffic, (b) slow traffic under the impact of multiple lanes blockage crashes}
    \label{fig:crash}
\end{figure}

As shown in Figure \ref{fig:MOTION-raw}, the test dataset primarily focuses on crashes and events where the model is expected to struggle. Figure \ref{fig:crash} discusses two typical types of crashes in the test dataset. Figure \ref{fig:crash}(a) illustrates crashes occurring within stop-and-go traffic. The performance of the reconstruction appears satisfactory, particularly when the bottleneck is evident in the conventional sensor data. However, it is clear that the fine-grained details do not align perfectly. Figure \ref{fig:crash}(b) depicts crashes that cause multiple lane blockages within the observation range, all occurring in slow-moving traffic. It is difficult to discern the wave details due to the lack of similar samples in the training data. These limitations necessitate more quantitative evaluation tools for assessing the generative results. Particularly, crash scenarios require further investigation with more attentions.

\subsection{Applicability and transferability}
\begin{figure}[H]
    \centering
    \includegraphics[width=0.75\linewidth]{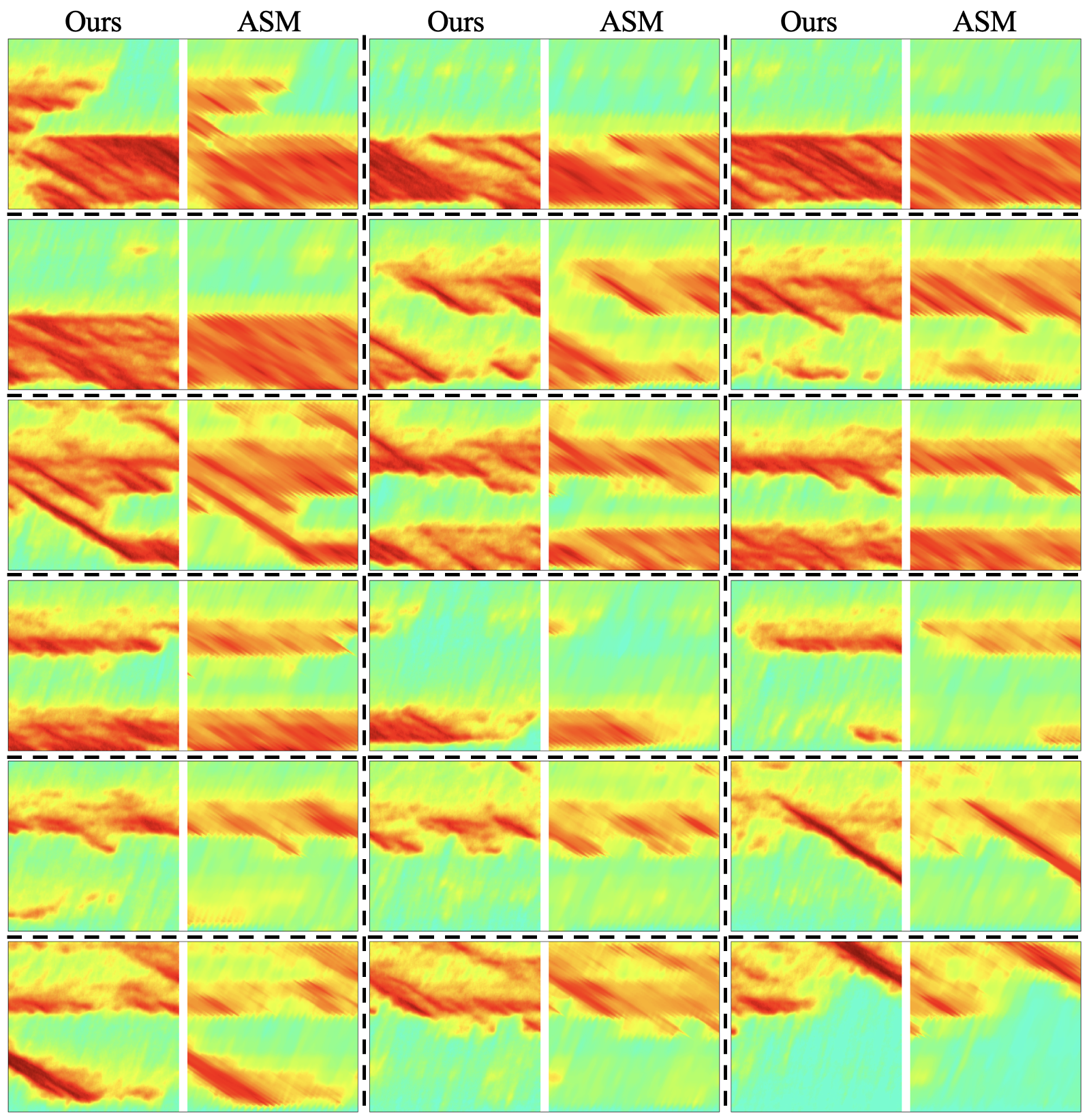}
    \caption{Transferability of the model to another freeway stretch (Mile Marker 54.5 to Mile Marker 58.5) on I-24 demonstrated on June 11, 2024. The model is trained on the I-24 MOTION stretch (Mile Marker 58.7 to Mile Marker 62.7).}
    \label{fig:transfer}
\end{figure}

The learned mapping from coarse-grained sensor data to fine-grained traffic patterns can be employed to reconstruct stop-and-go waves on other freeway segments, provided the following prerequisites are met:
\begin{enumerate}[label=(\roman*),noitemsep]
    \item The sensor network need have a similar spatial resolution and employ a 30-second temporal aggregation for sensors that measure speed, occupancy, and flow.
    \item The sensors need be properly configured to ensure data compatibility with the learned mapping.
\end{enumerate}
We demonstrate the trained model perform on an untrained segment in Figure~\ref{fig:transfer} on another road stretch of I-24 (Mile Marker 54.5 to Mile Marker 58.5). 
The proposed training process is transferable, and the model developed in this study can serve as an effective initialization for re-training on a different freeway segment, enabling fine-tuning \cite{moon2022fine} for new scenarios and different geographic locations.
\subsection{Challenges on data quality}
The quality of the data, particularly the accuracy of the ground truth fine-grained data, plays a critical role in model performance. While sporadic issues such as communication loss or data anomalies can often be mitigated through preprocessing in our upsampling module (ASM), systematic biases present a more persistent challenge. In the following, we provide a detailed discussion and evidence on the risks and impacts of data bias.
\begin{enumerate}[label=(\roman*),noitemsep]
    \item Biased training data, unbiased or biased input: It would clearly lead to biased generated outputs. This issue has been discussed in the generative modeling literature \cite{gao2025diffusionmeetsflow}, where the impact of training data bias on model generation distribution have been investigated \cite{kim2024training,hall2022systematic}.
    \item Unbiased training data, biased input: In our experiments, we introduce bias into the input data by adding Gaussian noise to the coarse-grained observations. Specifically, the noise is drawn from a normal distribution $\mathcal{N}(\mu, \sigma^2)$, where the mean $\mu$ and standard deviation $\sigma$ are independently sampled from the discrete set $\{0, 1, 2, 3, 4, 5\}$. This formulation allows us to evaluate the model performance on varying degrees and types of input bias.
    \begin{figure}[H]
        \centering
        \includegraphics[width=0.80\linewidth]{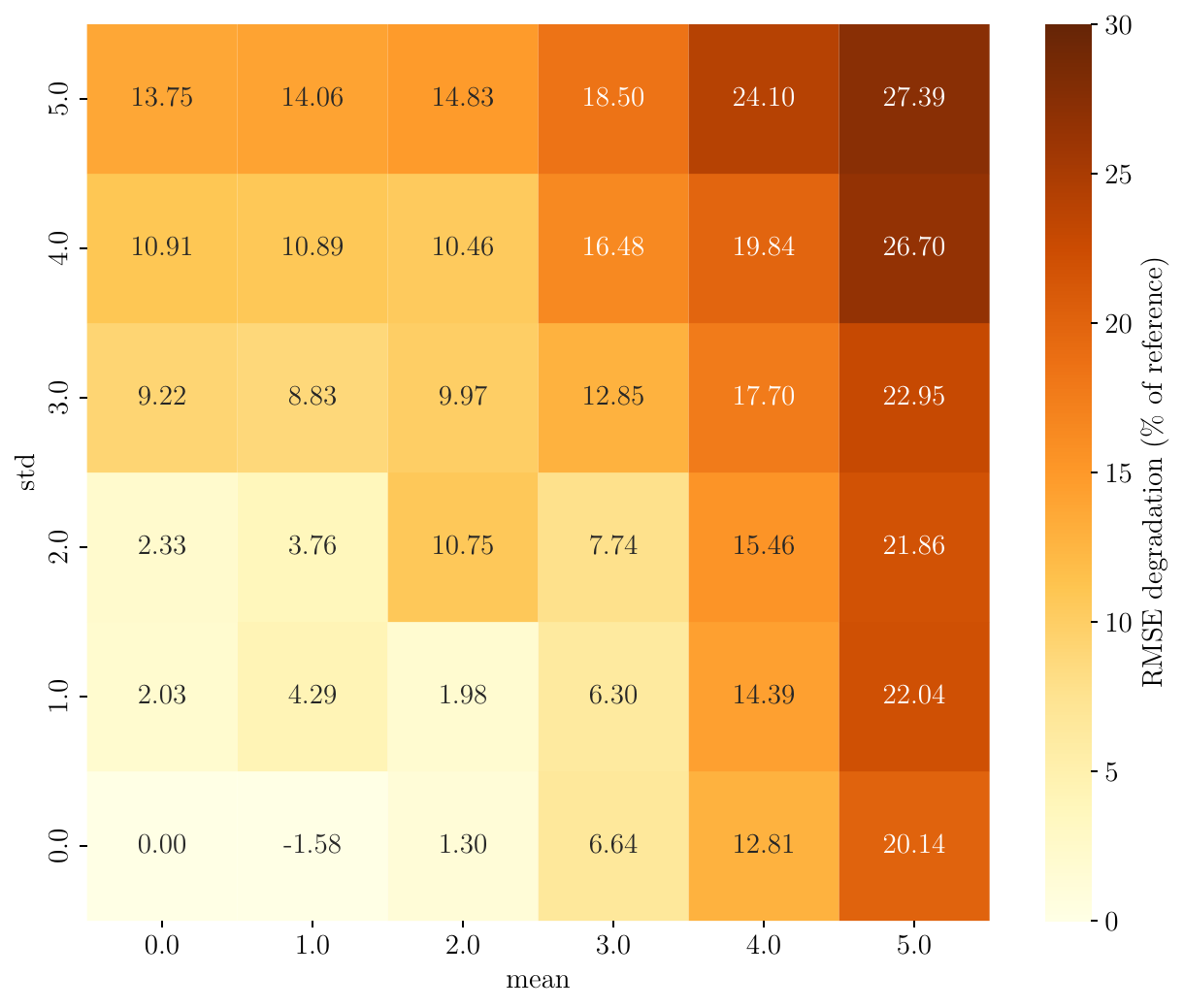}
        \caption{RMSE degradation under varying levels of Gaussian noise added to the raw data.}
        \label{fig:error-matrix}
    \end{figure}
    Figure~\ref{fig:error-matrix} demonstrates the degradation in model performance under varying levels of Gaussian noise, evaluated using data from June 3, 2024. The results indicate that the quality of input data impacts model performance. Specifically, bias in the mean value of the added noise (value bias) tends to have a more pronounced impact on RMSE degradation compared to random bias introduced through variations in the standard deviation. This suggests that systematic shifts in the data distribution are more detrimental to the model than random fluctuations of comparable scale. The overall RMSE degradation remains bounded within 30\% across the tested noise configurations.
    
    For sporadic issues such as communication loss or data anomalies, the upsampling component ASM is specifically designed to address these challenges, as illustrated in the bottom example of Figure~\ref{fig:val} and supported by prior studies \cite{treiber2002reconstructing,treiber2011reconstructing}.
\end{enumerate}

\section{Conclusion} \label{sec:conclusion}
This paper has demonstrated the potential of generative AI models to enhance the resolution of conventional traffic sensor data, thereby approximating the quality of high-fidelity observations. By employing a conditional diffusion denoising model, we are able to reconstruct fine-grained traffic speed fields from radar-based conventional sensors, effectively addressing the limitations associated with coarse-grain data resolution.

Our proposed approach is validated using a new dataset \texttt{WaveX} which provides a comprehensive ground for testing and comparing our model against various baselines. The results clearly show that our method outperforms the existing techniques across several evaluation metrics, notably in minimizing the Wasserstein Distance (WD) by 34.02\% on validation dataset and improving Root Mean Square Error (RMSE) and Mean Absolute Percentage Error (MAPE).  Despite spatio-temporal mismatches, our method significantly advances beyond current baseline methods on stop-and-go wave reconstructions by accurately capturing the wave patterns, leading to better estimations of travel times and stop-and-go wave speeds. However, limitations are also identified, particularly in scenarios involving crashes and complex traffic events, where fine-grained details do not align perfectly with high-fidelity data. 

In conclusion, our research contributes to the field of traffic data refinement by providing a viable method to improve the resolution of conventional traffic sensors. The open-sourcing of our dataset, trained model, and code enable further research and practical applications of generative AI in freeway operations.
\section*{Acknowledgments}
We are grateful for the thoughtful insights provided by our anonymous reviewers. This work was supported by the National Science Foundation (NSF) under Grant No. 2135579 (Work, Sprinkle), and 2111688 (Sprinkle) and the Tennessee Department of Transportation under Grant No. RES2023-20 and Grant No. OTH2023-01F-01. It is also supported by the U.S. Department of Transportation Dwight D. Eisenhower Fellowship program under Grant Agreement 693JJ32445065 (Richardson). We acknowledge the Tennessee Department of Transportation (TDOT) for providing the data used in this research. We are grateful to Prof. Yanbing Wang from Arizona State University for her contributions to the I-24 MOTION project. We also appreciate the insights for spatio-temporal modeling from Austin Coursey and Prof. Tyler Derr at Vanderbilt University. The views expressed herein do not necessarily reflect those of the U.S. Department of Transportation, the Tennessee Department of Transportation, or the United States Government. 
\bibliographystyle{ieeetr}
\bibliography{main}

\end{document}